\documentclass[prd,amsmath,amssymb,groupedaddress,letterpaper,twocolumn]{revtex4}
\usepackage{amsmath,amssymb,graphicx,bm} 
\usepackage{color}

\newcommand{\cF}{{\cal F}}
\newcommand{\cG}{{\cal G}}

\newcommand{\cL}{{\cal L}}

\newcommand{\cO}{{\cal O}}

\newcommand{\nn}{\nonumber}

\begin{document}

\title{Semi-classical wormholes are unstable}
 
\author{Roman~V.~Buniy}
\email{roman@uoregon.edu}
\author{Stephen~D.H.~Hsu}
\email{hsu@duende.uoregon.edu}

\affiliation{Institute of Theoretical Science, University of Oregon,
  Eugene OR 94703-5203}

\begin{abstract}
We show that Lorentzian (traversable) wormholes with semi-classical
spacetimes are unstable. Semi-classicality of the energy-momentum
tensor of the exotic matter used to stabilize the wormhole implies
localization of its wavefunction in phase space, leading to evolution
according to the classical equations of motion.  Previous results
related to violation of the NEC then require that the matter is
unstable to small perturbations.
\end{abstract}
 
\maketitle

\section{Introduction}\label{I}

The construction of a Lorentzian wormhole \cite{wormholes} or time
machine \cite{tm} requires violation of the null energy condition
(NEC):
\begin{equation}
T_{\mu \nu} n^\mu n^\nu \geq 0,
\end{equation}
where $T$ is the matter energy-momentum tensor and $n$ is any
null-vector ($n_\mu n^\mu = 0$). For example, in a wormhole the throat
geometry requires that converging null geodesics evolve into diverging
null geodesics (imagine the trajectories followed by light rays
traversing the wormhole). The Raychaudhuri equation for the expansion
$\theta$ of a hypersurface orthogonal null congruence is~\cite{he}
\begin{equation}
\frac{d\theta}{d \lambda} = -\tfrac{1}{2}\theta^2
-\sigma_{\mu\nu}\sigma^{\mu\nu}-R_{\mu \nu}n^\mu n^\nu,\label{theta}
\end{equation}
where $\lambda$ is an affine parameter, $n^\mu$ the tangent vector
field, and $\sigma^{\mu\nu}$ the shear. The first two terms on the
right-hand side of Eq.~(\ref{theta}) are negative, and after using the
Einstein equation to relate $R_{\mu \nu}$ and $T_{\mu\nu}$, we see
that $\theta$ can increase only if the NEC is violated.

In this paper we consider the exotic matter necessary to stabilize a
traversable wormhole throat (which may or may not be a time
machine). Our analysis focuses on constraining its properties, using
the fact that it violates the NEC and some additional assumptions
about the throat spacetime.

It is important to note that both the magnitude of $T_{\mu \nu}$ and
its violation of the NEC must be large in order to construct a useful
wormhole. Roughly, a wormhole with throat diameter of order $L$
requires energy-momentum densities of order $\sim l_P^{-2} L^{-2}$,
where $l_P$ is the Planck length. Taking $L$ to be of order meters
requires energy-momentum densities greater than fm$^{-4}$
~\cite{wormholes}. Similarly, it is easy to show using
Eq.~(\ref{theta}) that the time required to travel through the
wormhole is determined by the degree of violation of the NEC. This
travel time is $\gtrsim l_P^{-1} (- T_{\mu \nu} n^\mu
n^\nu)^{-\frac{1}{2}}$, where $T_{\mu \nu} n^\mu n^\nu < 0$. For
travel time of order 1 year, this requires $\vert T_{\mu \nu} n^\mu
n^\nu \vert \sim {\rm keV}^4$. Thus, the exotic matter under
consideration must exhibit large energy density and substantial
violation of the NEC.

We define two types of wormholes (or time machines constructed using
wormholes), henceforth referred to as ``devices'': those with
semi-classical spacetimes (type A) and those with strongly fluctuating
spacetimes (type B). Clearly, type A devices are preferable to type B
as they can by definition be controlled more precisely, and perhaps
pose less risk to users in their operation. Unfortunately, we will
show that all type A devices are unstable to small perturbations, so
time travel or travel via wormhole is possible only via an
intrinsically quantum device with a ``fuzzy'' spacetime. Essentially,
type A devices require exotic matter which behaves classically
(quantum effects are negligible in its dynamics), and classical matter
which violates the NEC is unstable \cite{nullbh}.

\begin{figure}[h!]
\includegraphics[width=8.5cm]{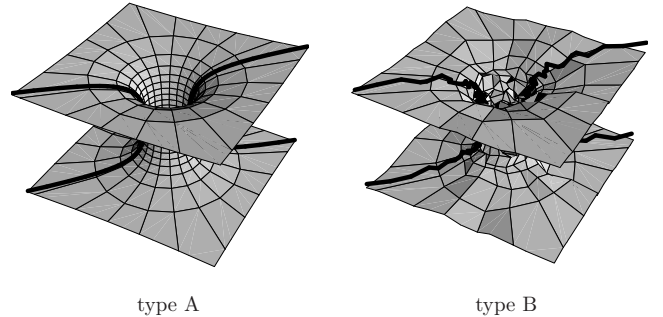}
\caption{Type A wormhole with semi-classical spacetime, and type B
wormhole with large spacetime fluctuations induced by quantum matter.}
\label{figurew}
\end{figure}

By a semi-classical state $\vert a \rangle$ we mean one whose
probability distribution is peaked about its central value in a
particular basis, defined by some operator $Z$. In this case,
\begin{equation}
\langle a \vert Z^n \vert a \rangle = \langle a \vert Z \vert a
\rangle^n
\label{Z}
\end{equation} 
up to small corrections (assuming $\langle a\vert a\rangle =1$), and
we say that $\vert a \rangle$ is semi-classical with respect to the
operator $Z$. This condition of semi-classicality is weaker than the
usual one, which requires that Eq.~(\ref{Z}) holds for {\it all}
operators simultaneously.

A state might, according to our definition, be semi-classical with
respect to $Z$ but not with respect to some other operator $Y$ (i.e.,
it might exhibit large fluctuations in measurements of $Y$). For
example, $\vert a \rangle$ might be an eigenstate of $Z$, while $[Y,Z]
\neq 0$ (e.g., let $Z$ be the momentum operator and $Y$ the position
operator). A central issue addressed in this paper is whether the time
evolution of a device which gives rise to a semi-classical, non-zero
$T_{\mu \nu}$ is well-approximated by the classical equations of
motion, or equivalently, whether it is simultaneously semi-classical
with respect to both the field operators $\phi$ and their conjugate
momenta $\pi$.

If the time evolution of the exotic matter in the device is
well-approximated by the classical equations of motion, we can apply a
previous result \cite{nullbh}, showing that in models built from gauge
fields, scalars and fermions, any classical solution is unstable to
small perturbations if it violates the NEC. Specifically, in
Ref.~\cite{nullbh} it was shown that the kinetic term of the effective
Hamiltonian governing small fluctuations about a classical solution
must exhibit a negative eigenvalue if the NEC is violated. Therefore,
there exist infinitesimal perturbations involving local spatial
gradients which lower the energy of the solution.

\section{Spacetime fluctuations}\label{SF}
 
In this section we discuss spacetime fluctuations \cite{spacetime},
and argue that type A devices are necessarily semi-classical with
respect to the matter energy-momentum tensor $T_{\mu \nu}$.

It is useful to distinguish between ``active'' and ``passive'' metric
fluctuations. Active fluctuations are those involving quantum effects
due to the graviton itself; these are negligible as long as the
spacetime curvature is small compared to the Planck scale. We focus on
passive fluctuations which are induced on the metric through the
Einstein equations $G_{\mu \nu} = 8 \pi T_{\mu \nu}$ by fluctuations
in $T_{\mu \nu}$. Clearly, in order for the spacetime to be
semi-classical, the matter state must be semi-classical with respect
to $T_{\mu \nu}$. 

Because of ultraviolet divergences, the property of semi-classicality
is scale-dependent. On sufficiently small length scales the passive
fluctuations become arbitrarily large. Operator expectations as in
Eq.~(\ref{Z}) require regularization; for example, if $Z = T_{\mu
\nu}$, it involves products of fields at the same point in
spacetime. We assume a regulator such as point-splitting or smearing
of the fields, controlled by a length scale $\sigma$. Short-distance
contributions to operator averages are then of order $\sigma^{-1}$ to
the appropriate power, leading, e.g., to fluctuations
\begin{equation}
\Delta T_{\mu \nu} \equiv \left( \langle T_{\mu \nu} T_{\mu \nu}
\rangle - \langle T_{\mu \nu} \rangle^2 \right)^{\frac{1}{2}} \sim
\sigma^{-4}
\end{equation} 
(no summation over indices). To investigate the effects of these
fluctuations, we take $\sigma$ to be some characteristic length scale
associated with the spacetime or relevant experimental probes, such as
test particles.

Minkowski spacetime is type A on length scales $\sigma > l_P$ (where $l_P$
is the Planck length), even though $\langle 0 \vert T_{\mu\nu} \vert 0
\rangle = 0$, so that $\Delta T_{\mu\nu} \sim \sigma^{-4} > \langle 0
\vert T_{\mu\nu} \vert 0 \rangle$ for any $\sigma$. This is because
the effect of these fluctuations on the metric $g$ is suppressed by
$l_P$, as can be seen from the Einstein equations: $\Delta G_{\mu\nu}
\sim l_P^2 \Delta T_{\mu\nu} \sim l_P^2 \sigma^{-4}$.  Since
$G_{\mu\nu}$ is constructed of objects with two derivatives of $g$, we
see that deviations of $g$ from flat space on length scales $\sigma$
are of order $l_P^2 \sigma^{-2} \ll 1$.  Borgman and Ford have studied
the effect of vacuum energy fluctuations on null geodesics in
Minkowski space, and found them to be undetectable \cite{BF}.

Now consider a device whose spacetime is nontrivial --- i.e., not flat
space. If we take $\sigma$ to be some characteristic length scale
associated with the device spacetime, the regulated short-distance
fluctuations are small compared to the central values of the energy
momentum tensor: $\langle a \vert T_{\mu\nu} \vert a \rangle \gg
\sigma^{-4}$. In fact, from the Einstein equations we expect $\langle
a \vert T_{\mu\nu} \vert a \rangle \sim l_{P}^{-2} \sigma^{-2}$. In
other words, the energy and momentum densities required to warp
spacetime on length scale $\sigma$ are much larger than
$\sigma^{-4}$. (As mentioned, in the wormholes studied by Morris and
Thorne in Ref.~\cite{wormholes}, for a throat size of order meters the
required energy-momentum densities are of order fm$^{-4}$, provided
perhaps by a cosmic string or similar object.) Then, the question of
whether the spacetime is semi-classical depends not on the
short-distance singularities in the operator products, but rather on
whether the wavefunction of $\vert a \rangle$ is highly peaked about
its central value on length scales of $\sigma$ or larger. \emph{The
classicality of the spacetime is determined by the classicality of the
energy-momentum tensor in the matter state $\vert a \rangle$ .}

As an example, consider the Casimir vacuum $\vert c \rangle$, which
describes the interior of a cavity of size $L$. The NEC is violated in
the Casimir state, which has negative (renormalized) energy density:
$\langle c \vert T_{00} \vert c \rangle \sim - 1/L^4$. However, since
the fluctuations $\Delta T$ are of order $1/\sigma^4$, they are larger
than the central value even if we choose the largest possible length
scale $\sigma \sim L$; the Casimir state never leads to a
semi-classical energy-momentum tensor. Further, the Casimir state is
problematic for construction of an interesting device, since the
curvatures induced are only of order $l_P^2 L^{-4} \ll
L^{-2}$. Spacetime within the cavity deviates from flat space only
very slightly, and in an intrinsically fuzzy (highly fluctuating)
manner.

\section{Energy-momentum tensor and phase space}\label{EMT}

In this section we show that semi-classicality with respect to $T_{\mu
\nu}$ implies semi-classicality with respect to the fields $\phi$ and
their conjugate momenta $\pi$. Hence the wavefunction is localized in
phase space, and the quantum evolution is closely approximated by the
classical evolution. The proof proceeds in two steps. First, we note
that at least two components of $T_{\mu \nu}$ do not commute with each
other. Then, we derive a relation between the uncertainties in two
such non-commuting operators and arbitrary functions of them. In the
case of interest, the original pair is two non-commuting components of
$T_{\mu \nu}$ and the functions map these operators into $\phi$ and
$\pi$. The relation shows that small uncertainties in $T_{\mu\nu}$
imply small uncertainties in the phase space variables.

First, using the translation operator, it is easy to see that $F_{\mu
\nu \rho \sigma} (x,x') = [T_{\mu\nu}(t,x),T_{\rho \sigma}(t,x')]$ is
related to $F_{\mu \nu \rho \sigma}(x-y,x'-y)$ by a unitary
transformation.  Therefore, if for some $x$, $F_{\mu \nu \rho \sigma}
(x,x')=0$ for all $x'$, then $F_{\mu \nu \rho \sigma} (x,x')=0$
identically. However, if that were the case we would have, in
particular, $[T_{\mu\nu}(t,x),T_{00}(t,x')]=0$ for all $x$ and
$x'$. Integrating over $x'$, we find that $T_{\mu\nu}(t,x)$ does not
depend on time, which is possible only if $\cL=\text{const}$. Thus, in
any non-trivial model, for any $x$ there always exists an $x'$ for
which $F_{\mu \nu 00} (x,x') \neq 0$. In fact, $F_{\mu \nu \rho
\sigma} (x,x')$ has support only for $x$ near $x'$, since it is
proportional to the delta function $\delta (x-x')$ or its spatial
derivatives; this follows from the canonical commutation relation
$[\phi(t,x), \pi(t,x')] = i \delta (x-x')$. Thus, there are at least
two non-commuting components of the energy-momentum tensor.

Next, suppose we have a pair of non-commuting operators $A$ and $B$,
and consider a new operator which is an arbitrary function of them,
$F=\cF(A,B)$. The only limitation imposed on the function $\cF$ is
that it can be expanded in a power series of its arguments. (Note that
if $A$ and $B$ commute we cannot necessarily express an arbitrary $F$
as a local function of them. For example, let $A = x$, $B = x^2$ and
$F = p$.) It should be clear that if both $A$ and $B$ are
semi-classical, than $F$ is semi-classical too. Indeed, let us expand
the operator $F$ around its classical value
$\bar{\cF}=\cF(\bar{A},\bar{B})$:
\begin{eqnarray}
F&=&\bar{\cF}+\bar{\cF}_{\bar{A}}(A-\bar{A})
+\bar{\cF}_{\bar{B}}(B-\bar{B})\nn\\
&+&\tfrac{1}{2}\bar{\cF}_{\bar{A}\bar{A}}(A-\bar{A})^2
+\tfrac{1}{2}\bar{\cF}_{\bar{B}\bar{B}}(B-\bar{B})^2\nn\\
&+&\tfrac{1}{2}\bar{\cF}_{\bar{A}\bar{B}}[A-\bar{A},B-\bar{B}]
+\cO(\Delta^3),
\end{eqnarray}
where $\Delta$ is of order $\Delta A$ or $\Delta B$. The fluctuation
$(\Delta F)^2=\langle F^2 \rangle -\langle F \rangle^2$ then is
\begin{eqnarray}
(\Delta F)^2 &=& \bar{\cF}_{\bar{A}}^2(\Delta A)^2
+\bar{\cF}_{\bar{B}}^2(\Delta B)^2\nn\\
&+&\bar{\cF}_{\bar{A}}\bar{\cF}_{\bar{B}}(\langle AB \rangle +\langle
BA \rangle-2\bar{A}\bar{B}) +\cO(\Delta^3).\label{Delta.F}
\end{eqnarray} 
For semi-classical $A$ and $B$, the third term on the right hand side
of Eq.~(\ref{Delta.F}) is as small as the first two terms. Thus $F$ is
semi-classical if $A$ and $B$ are.

If there is another operator $G=\cG(A,B)$, we also expand it around
its classical value $\bar{\cG}=\cG(\bar{A},\bar{B})$ and find the
commutator
\begin{eqnarray}
[F,G]=(\bar{\cF}_{\bar{A}}\bar{\cG}_{\bar{B}}
-\bar{\cF}_{\bar{B}}\bar{\cG}_{\bar{A}})[A,B]
+\cO(\Delta^3).\label{Jacobian}
\end{eqnarray} 
Alternatively, in the semi-classical limit the commutator $i[F,G]$
turns into the Poisson bracket $\hbar\{\bar{\cF},\bar{\cG}\}$. Since
the brackets $\{\bar{\cF},\bar{\cG}\}$ and $\{\bar{A},\bar{B}\}$ are
related via the Jacobian
$J=\partial(\bar{\cF},\bar{\cG})/\partial(\bar{A},\bar{B})$, we arrive
at Eq.~(\ref{Jacobian}) again.

Fluctuations of two semi-classical operators $A$ and $B$ satisfy the
relation $\Delta A\Delta B \sim \hbar \langle [A,B] \rangle$. Since
$F$ and $G$ are semi-classical too, writing the similar relation for
them and combining the result with Eq.~(\ref{Jacobian}), we arrive at
\begin{eqnarray}
\Delta F\Delta G\sim J\Delta A\Delta B. 
\end{eqnarray}

\begin{figure}[h!]
\includegraphics[width=8.5cm]{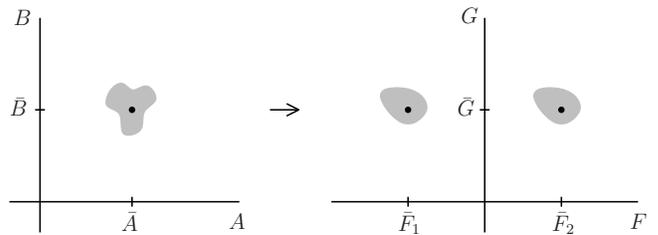}
\caption{The diagram illustrates the relation between localization in
  the $(A,B)$ and $(F,G)$ spaces, up to possible degeneracy.}
\label{figure}
\end{figure}

To complete our proof, we identify the operators $A,B$ with two
non-commuting components of $T_{\mu\nu}$, and the operators $F,G$ with
$\phi,\pi$. A state which has semi-classical energy-momentum tensor
then must be localized in phase space, and therefore evolve according
to the classical equations of motion. Note that if either of the
commutators in Eq.~(\ref{Jacobian}) vanish, the Jacobian $J$ would be
either zero or divergent, and localization in $(A,B)$ space would not
imply localization in $(F,G)$ space, or vice versa.

It is possible that degeneracies (either discrete or continuous) lead
to multiple small regions in the $(F,G)$ space which are mapped onto
the same small region of the $(A,B)$ space (see
Fig.~(\ref{figure})). In other words, there may be inequivalent
configurations in phase space with the same energy-momentum tensor. In
such a case, the preceding equations should be modified by including
the sums (or integrals, in the continuous case) over these
regions. For example, if two states $\vert a_1 \rangle$ and $\vert a_2
\rangle$ with distinct values of $\langle \phi \rangle$ or $\langle
\pi \rangle$ lead to the same $\langle T_{\mu\nu} \rangle$, we can
obtain a semi-classical spacetime from a superposition of the
two. Nevertheless, {\it each} component of the superposition is
localized (behaves semi-classically) and will separately exhibit an
instability.

\section{Discussion}

Classical systems which violate the NEC are unstable \cite{nullbh} in
a particularly violent way: they can lower their energy by increasing
local spatial field gradients. Therefore, the exotic matter used to
stabilize a wormhole throat must be quantum mechanical in nature; in
other words, quantum effects must play an important role in its
dynamics and time evolution.  However, it is undesirable for a device
to have a strongly fluctuating (non-semi-classical) spacetime. Such a
device would presumably behave unpredictably and might transport its
payload to an undesirable time or place. We have shown that
semi-classicality of the spacetime is a strong enough condition to
imply phase space localization of the wavefunction of the stabilizing
matter. This means the time evolution of type A devices will be
semi-classical and well-approximated by the classical equations of
motion. Such devices are unstable in any region where the NEC is
violated. Wormholes cannot be both predictable and stable.


\begin{center}
\textbf{Acknowledgments}
\end{center}

This work was supported by
the Department of Energy under DE-FG06-85ER40224.

\end{document}